\documentclass[11pt,twoside,a4paper,cr,final]{cms-tdr}
\def\svnVersion{62658:62698}
\pdfoutput=1
\def\cmsMessage{Presented at Moriond/EW: Rencontres de Moriond on ``EW Interactions and Unified Theories"}
\newcommand{\tauhad}{\tau_{\mathrm{had}}}
\newcommand{\THELUMI}{{\ensuremath{18.4\pm 0.7~{\mathrm{pb}}^{-1}}}}
\newcommand{\RHT}{R_{\mathrm{HT}}}
\begin{document}

\hyphenation{had-ron-i-za-tion}
\hyphenation{cal-or-i-me-ter}
\hyphenation{de-vices}

\cmsNoteHeader{-2011/061} 
\title{Observation of $W\rightarrow\tau\nu$ Production\\
in pp Collisions at $\sqrt s = 7 $ TeV}

\author[cms]{Abdollah Mohammadi for the CMS Collaboration}


\address[cms]{School of Particles and Accelerators, Institute for Research in
Fundamental Sciences (IPM), P.O.Box 19395-5531, Tehran, Iran;  also at: \\
Shiraz University, Physics Department, 71454, Shiraz, Iran }


\abstract{
   The production of W bosons decaying into a tau lepton and a neutrino 
with the tau lepton decaying hadronically has been observed in LHC 
pp collisions at $\sqrt{s} = 7$~TeV with the CMS detector. 
The selection criteria provide a statistically significant signal
on the top of QCD multi-jet and electroweak
backgrounds. A data-driven method for the estimation
of the QCD multi-jet background has been employed.}

\hypersetup{%
pdfauthor={CMS Collaboration},%
pdftitle={CMS Paper Template 2006 LaTeX/PdfLaTeX version},%
pdfsubject={CMS},%
pdfkeywords={CMS, physics, software, computing}}

\maketitle 

\section{Introduction}
Tau leptons serve as an important probe for many new physics processes at the LHC. 
Among others, experimental signatures that involve decays to tau leptons are 
crucial for searches of a light Higgs boson, Supersymmetry or extra dimensions.

Tau leptons can decay either leptonically via $\tau \rightarrow l \nu \bar\nu$ (l=e or mu, branching fraction is  36\%) or
into a hadronic jet and one tau-neutrino. Hadronic decay modes  ($\tauhad$) produce a highly 
collimated tau-jet signature, characterized by a low particle multiplicity 
that allows their separation from QCD-jets.

In the framework of the standard model, tau leptons are mostly produced in decays of 
electroweak vector bosons: $Z\rightarrow\tau^{+}\tau^{-}$ and
$W^{\pm}\rightarrow\tau^{\pm}\nu$. These processes have relatively large 
cross sections and are among the largest sources of tau leptons at LHC. 
The $W\rightarrow\tau^{\pm}\nu$ channel benefits from a large production 
cross section, exceeding the production rate of $Z\rightarrow\tau^{+}\tau^{-}$ 
by nearly an order of magnitude. However, the experimental signature of a 
single tau-jet and undetected neutrino is challenging, requiring a good 
understanding of the tau identification and missing transverse energy ($\MET$).

The study of $W\rightarrow\tau\nu$ production in the $\tauhad\nu$ final 
state is an important calibration sample for understanding
tau identification and reconstruction. Also, $W^{\pm}\rightarrow\tau^{\pm}\nu$ production has to be well understood 
as a test of the standard model and as a measure of important background process
in several searches for new physics. In particular, it is the major background in
the search for the charged Higgs boson in the $\tau\nu$ final state.

This study of $W\rightarrow\tauhad\nu$ production has been conducted using 
$\THELUMI$ of collision data from the 2010 LHC run at $\sqrt{s} = 7$~TeV 
recorded with the CMS detector.  See Ref.~\cite{ZTAUTAU} for a measurement
of the cross section for $Z\rightarrow\tau^{+}\tau^{-}$ production 
including tau-leptons reconstructed in the $\tauhad$ final state.

\section{Physics objects reconstruction}
The particle flow (PF) reconstruction algorithm implemented at CMS~\cite{CMS_PAS_2009-001} 
is used for identification of jets, muons, electrons, taus and $\MET$.
The PF technique utilizes the information from the whole event, aiming 
to provide a global event description at the level of individually 
reconstructed particles. Firstly, all tracks and energy clusters are 
reconstructed in each sub-detector. Next, all the candidates are 
associated in an optimal combination to one or more of these sub-detector 
signals, if they are compatible with the physics properties of each particle, 
and reconstructed in the event. The final set of particles (charged hadrons,
neutral hadrons, photons, electrons and muons) is used to derive 
composite physics objects such as $\tauhad$, jets and $\MET$. 
The PF jets are clustered using the anti-$k_T$ jet clustering 
algorithm~\cite{jet14} with distance parameter $R = 0.5$.

Typically, $\tauhad$ is a highly-collimated jet comprising one or 
three charged mesons (predominantly $\pi^{\pm}$) and possibly one or two
neutral pions always decaying via $\pi^{0} \to \gamma \gamma$. 
The identification of $\tauhad$ from W~boson decays requires a robust 
algorithm and an efficient set of selection criteria, as it is one of 
the main discriminators against large QCD jet background. 

The $\tauhad$ identification algorithm used here 
is known as the Hadrons Plus Strips Algorithm (HPS)~\cite{CMS_PFT-10-004}.
HPS starts from a high-$p_T$ charged hadron and combines it 
with other nearby charged or neutral hadrons to reconstruct
$\tau $ decay modes. The identification of $\pi^0$s is 
enhanced by clustering the PF electrons and photons in 
"strips" along the bending plane to take into account 
possible broadening of calorimeter signatures because of photon conversions.

\section{Event selection}

\begin{figure}[hbtp]
  \begin{center}
    	\includegraphics[width=0.45\textwidth]{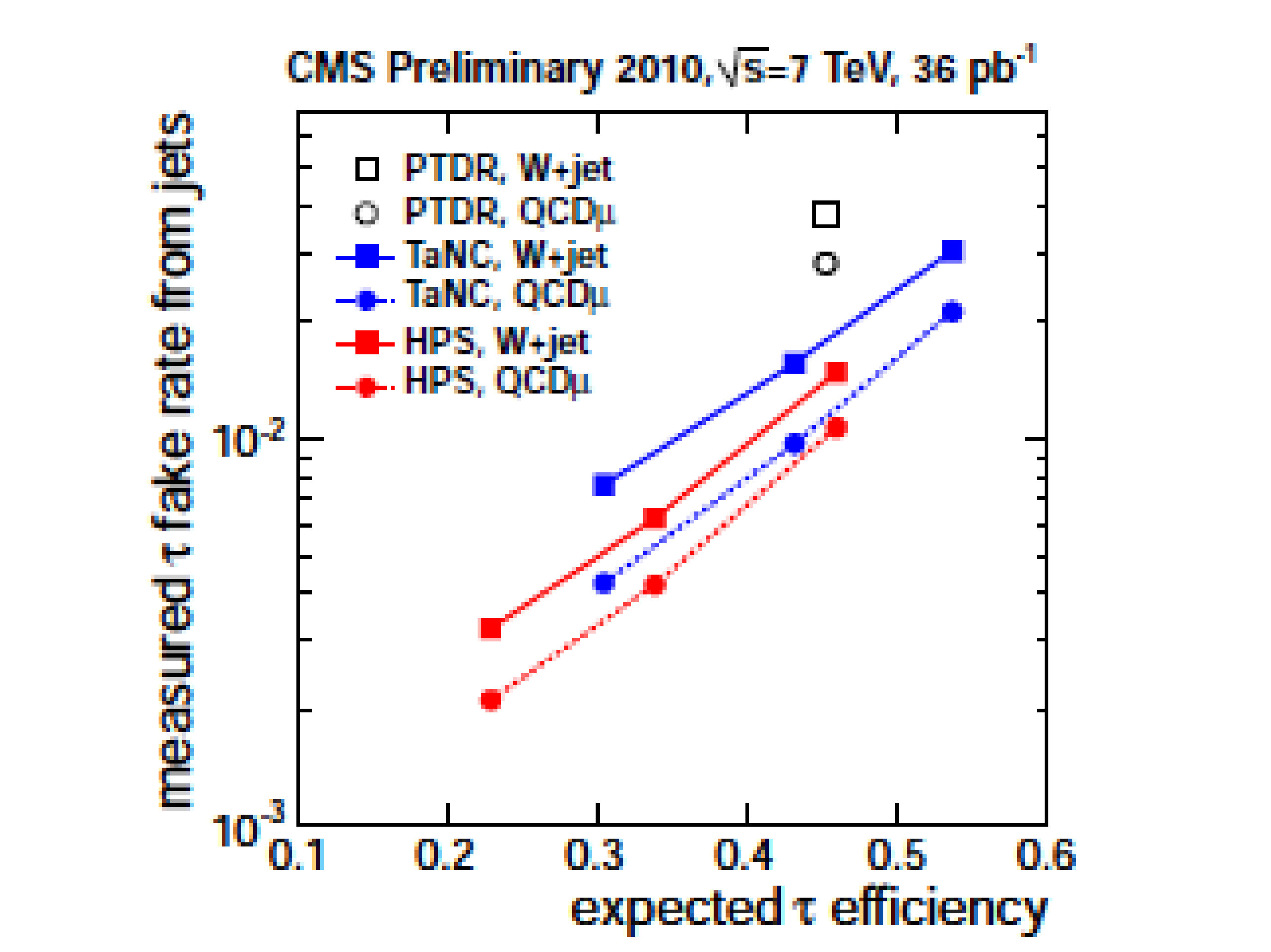}
	\caption{ \label{TauId}
The measured fake rate as a function of efficiency evaluated using simulation for all working
points for QCD m-enriched and W data samples. The PTDR points represent results of the
fixed cone algorithm based on the PF taus}
  \end{center}
\end{figure}

The following list of offline selection criteria is applied for the final event selection:
\begin{itemize}
\item
There must be at least one HPS $\tauhad$ candidate with 
$p_T > 30$~GeV and $|\eta| < 2.3$, and the leading track 
in the $\tauhad$ candidate must have $ p_T > 15$~GeV.
 Three different working points (Figure~\ref{TauId}) for the isolation has been defined~\cite{TAU_11_001}. The definition of the medium, which has been used in this analyis follows as: there must be no PF charged hadron or photon candidates with $p_T > 0.8$~ GeV within an isolation cone of size $\Delta R = 0.5$, (Those candidates which are associated to the tau decay signature are excluded.)
\item
Several cuts has been applied in order to rejects those electron and muons which fake taus. Furthermore 
we also veto the events which include good electron or muon. This cut supresses the W+Jet events where W decays either to muon or electron and jet fakes tau.

\item 
We require $\MET > 35$~GeV and we consider PF jets in an event with $p_T > 15$~GeV and 
$|\eta| < 3$, and compute the ratio, $\RHT$, of the $p_T$
of the $\tauhad$ candidate to the sum of the $p_T$ of the PF jets.
We require $\RHT > 0.65$.
\\
\\
Further details about event selection can be found in elsewhere~\cite{EWK_11_002}.
\end{itemize}

\section{QCD Background Estimation}
\par
QCD events are the dominant background contribution to the
final event sample.  This background cannot be reliably
estimated from simulation, so a data-driven method is used.

In the so-called ``ABCD method,'' four regions are designated in
a phase space defined by $\MET$ and $\RHT$.  We start with an
event sample obtained with no cuts on $\MET$ and $\RHT$,
and then divide it into four subsamples as follows
\begin{itemize}
\item region A where $\RHT > 0.65$ and $\MET > 35$ GeV. 
This region is dominated by signal; we want to account for
QCD background here.
\item region B where $\RHT  > 0.65$ and $\MET < 35$ GeV
\item region C where $\RHT < 0.65$ and $\MET < 35$ GeV
\item region D where $\RHT < 0.65$ and $\MET > 35$ GeV.
\end{itemize}
In order to apply this method, we must assume that the
event subsamples in regions B, C and D are dominated
by QCD events, and there is a low statistical
correlation between~$\RHT$ and~$\MET$.  All other backgrounds have been neglected and no corrections
have been applied due to the signal contribution in the B, C and D regions.

\begin{figure}[hbtp]
  \begin{center}
    	\includegraphics[width=0.7\textwidth]{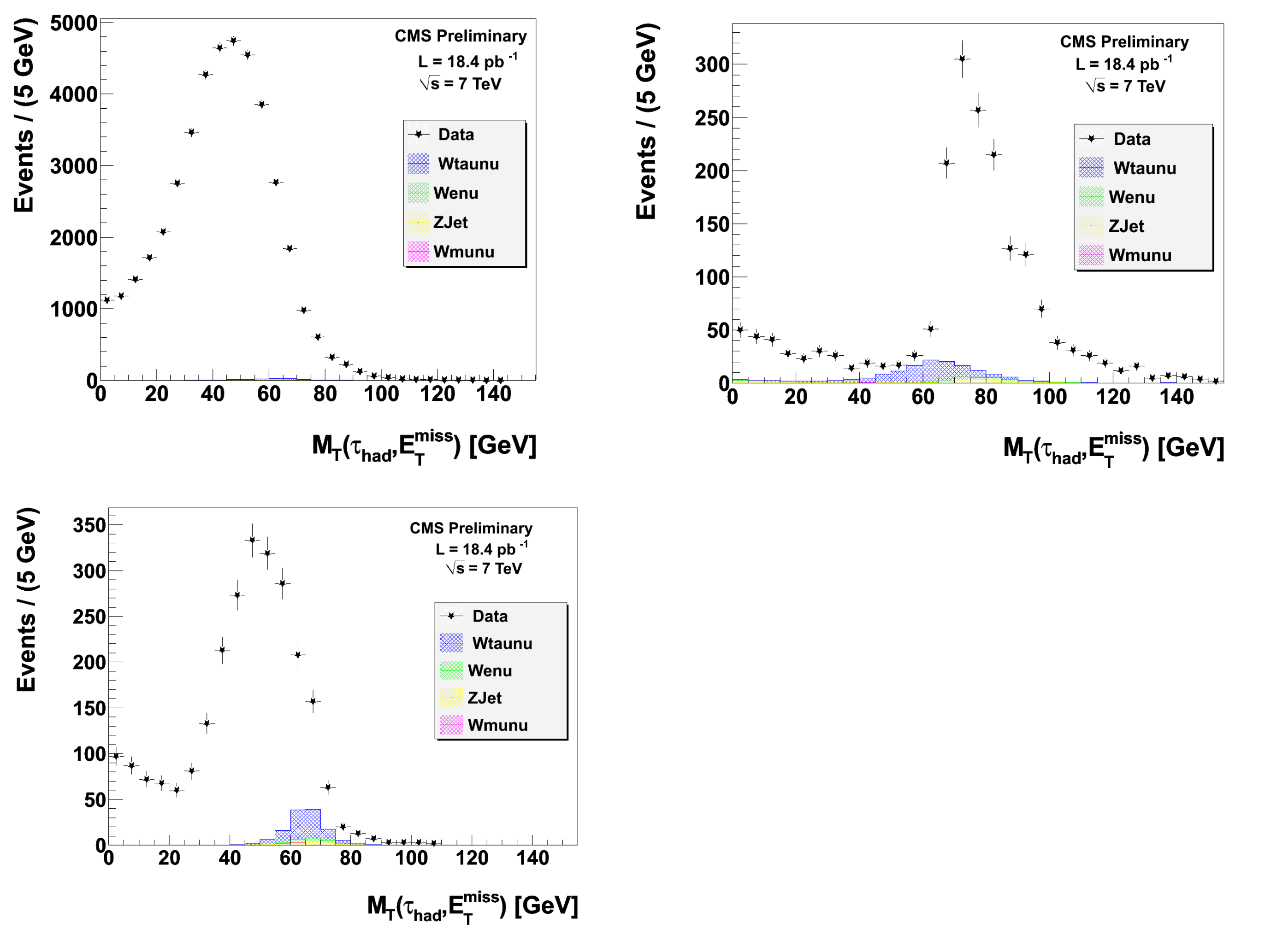}
\caption{ \label{ABCDTMass}
Transverse mass distributions of the $\tauhad$ candidate and $\MET$
for the four designated regions in phase space: 
Region~B (bottom left) where $\RHT  > 0.65$ and $\MET < 35$~GeV,
Region~C (upper left) where $\RHT < 0.65$ and $\MET < 35$~GeV,
and Region~D (upper right) where $\RHT < 0.65$ and $\MET > 35$~GeV.  
The points represent the data.  Simulated signal and electroweak
backgrounds are represented by the filled histograms.}
  \end{center}
\end{figure}

Figure~\ref{ABCDTMass} illustrates the transverse mass 
distributions of $\tauhad$ candidates and $\MET$ in regions
B, C and D. One sees that indeed these regions are dominated
by QCD background.  The signal contribution is less than~1\%
in region C, and less than 5\% and 10\% in regions B and D,
respectively. It has been shown~\cite{EWK_11_002} that the level of correlation between $\RHT$ and $\MET$ is sufficiently low to yield accurate background estimation using the ABCD method



We estimate the yield of QCD background events in the signal region~A
from the numbers of events observed in the other regions.
Specifically, we assume $N_A = (N_D\times N_B)/N_C$,
and obtain $N_A = 109\pm 6$~events, where the uncertainty
is statistical only.

\section{Results}

After all selections, the expected yield of $W \rightarrow \tau \nu$ events as well as electroweak background contributions are estimated using
simulation while the QCD multi-jet background is estimated from the ABCD method described above. With
the selections used in this analysis, number of signal event  is estimated to be $174 \pm 3$ (stat), the number of
electroweak backgrounds (dominated by  $W \rightarrow e \nu$ ) is estimated as $46 \pm 2 $ (stat) and the QCD multijet contribution is $109 \pm 6$ (stat). The number of events observed in data is $372$.

It should be mentioned that we have not yet assessed systematic uncertainties on the
background predictions or on the signal efficiency.

\begin{figure}[hbtp]
  \begin{center}
    \includegraphics[angle= 90,width=0.35\textwidth]{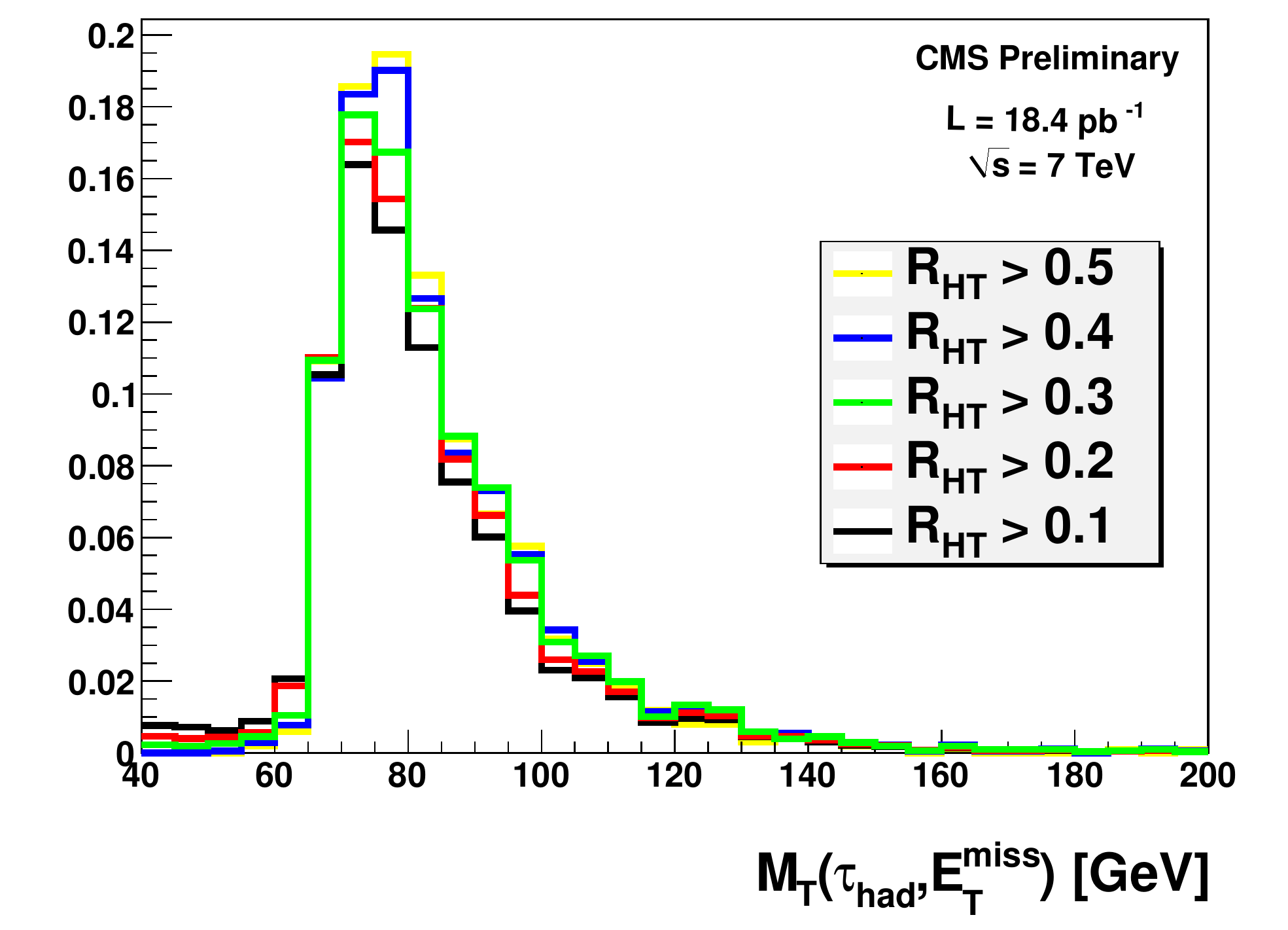}\hspace{0.5cm}
    \hspace{0.5cm}
    \includegraphics[angle= 90,width=0.35\textwidth]{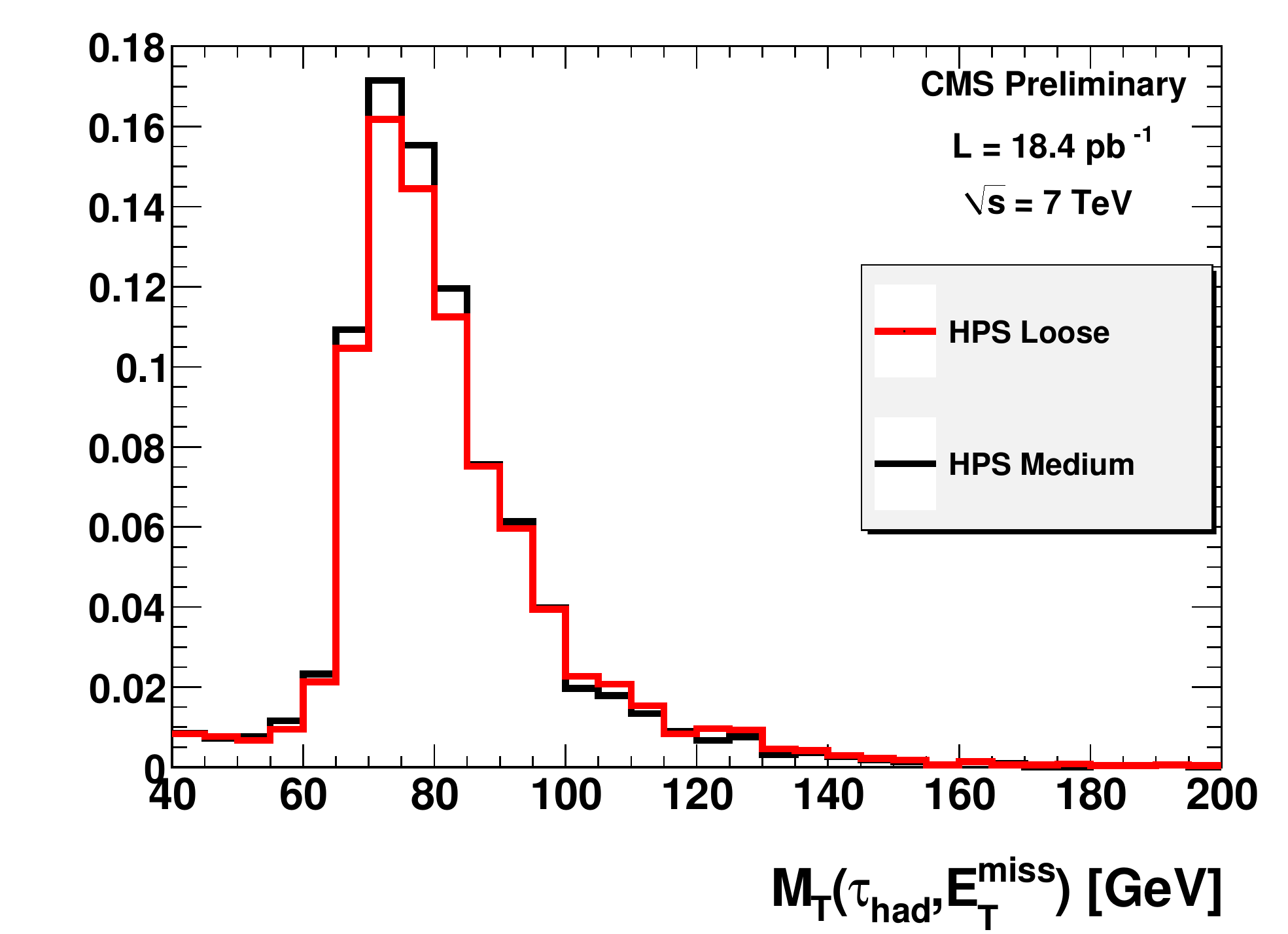}\hspace{0.5cm}
	\caption{Effect of changing the $\RHT$ and isolation criteria on the shape of Transverse Mass of $\tauhad$ and $\MET$.
}    
	\hspace{0.5cm}
    \label{Shape}
  \end{center}
\end{figure}

The shape of the transverse mass distribution for QCD multi-jet
events is estimated using a data-driven method. The strategy is to relax some cuts to move
into a region where QCD is dominant, and normalize this shape to the number 
of QCD events estimated with the ABCD method. 
Figure~\ref{Shape} shows that when changing the isolation criterion 
or the $\RHT$ cut (from 0.1 to 0.5),
the QCD shape does not change drastically.  We decided to use a working 
point where the contribution of electroweak processes and signal events 
under the mass peak is reduced to 15\%, loosening the cut on $\RHT$
from 0.65 to 0.3 and using a looser isolation requirement.

\begin{figure}[hbtp]
  \begin{center}
    \includegraphics[angle = 90, width=0.5\textwidth]{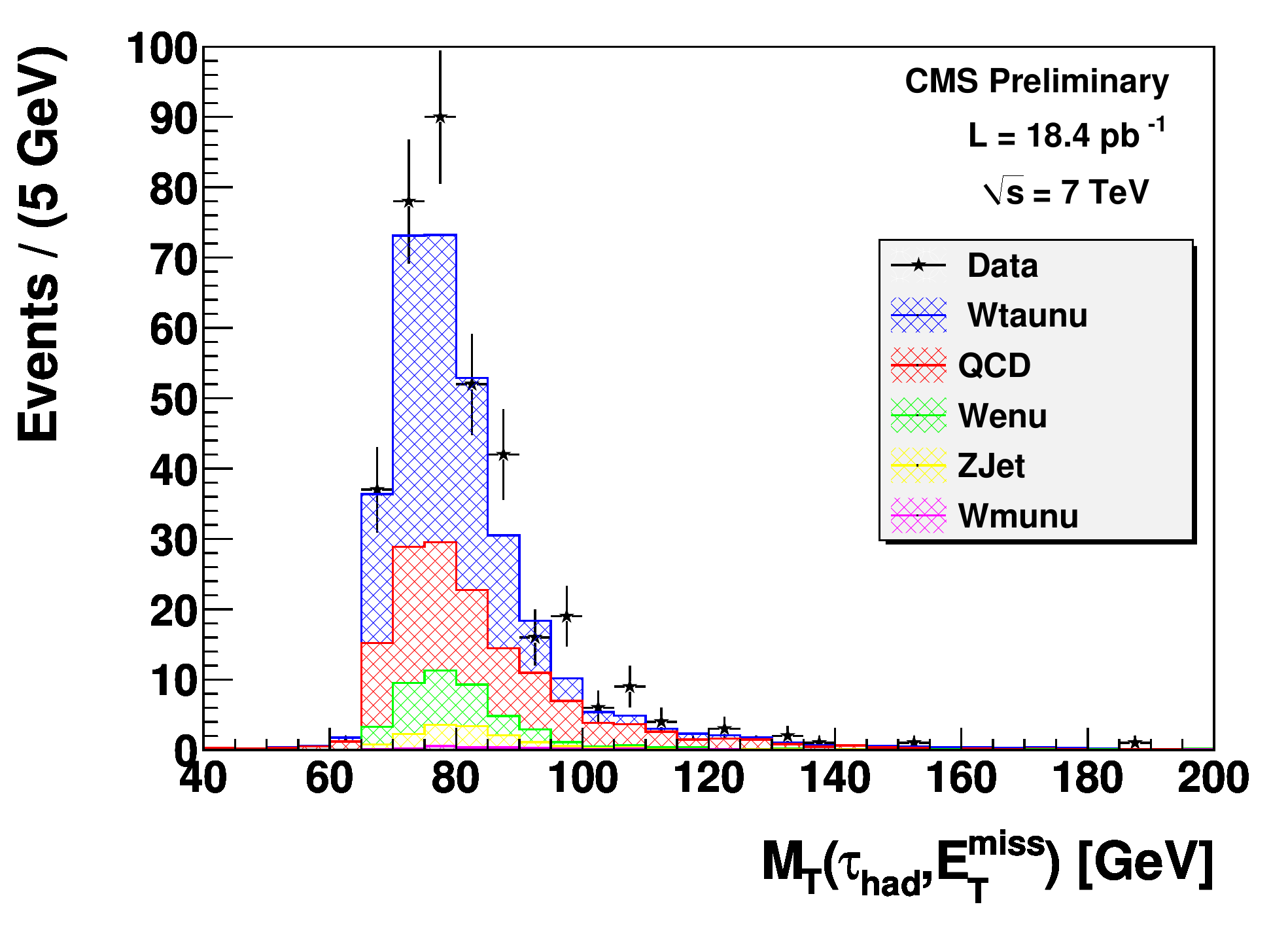}\hspace{1cm}
    \hspace{1cm}
    \caption{Transverse Mass of $\tauhad$ and $\MET$ after all cuts}
    \label{TMass}
  \end{center}
\end{figure}

Figure~\ref{TMass} shows the transverse mass distribution for the
final event sample, with the data-driven estimate of the QCD 
contamination. 

\section{Summary}
We have statistically significant signal for $W\rightarrow\tauhad\nu_{\tau}$
with the $\tau$-lepton reconstructed in its hadronic decay modes,
using $\THELUMI$ of data collected by the CMS Collaboration.

\section*{Acknowledgments}
I would like to thank the CMS collaboration especially, L.Rebana, A.Nikitenko, M.Khakzad, M.V.Acosta,  G.Bagliesi, M.Bachtis and  A.N. Safonov for their helpful comments and cooperations. 


\end{document}